\documentclass[aps,pre,twocolumn,superscriptaddress,floatfix,longbibliography]{revtex4-1}

\usepackage[dvips]{graphicx}
\usepackage{latexsym}
\usepackage{amsmath}
\usepackage{amsthm}
\usepackage{amsfonts}
\usepackage{amssymb}
\usepackage{mathptmx}
\usepackage{subfigure}

\usepackage{xcolor}
\definecolor{darkblue}{rgb}{0,0.02,0.45}
\usepackage[colorlinks=true,citecolor=darkblue,urlcolor=darkblue]{hyperref} 

\usepackage[all]{hypcap}
\usepackage{gensymb}

\definecolor{cream}{RGB}{222,217,201}     

\begin{document}

\title{Lattice dynamics and spin excitations in the metal-organic framework [CH$_3$NH$_3$][Co(HCOO)$_3$]}

\author{Lei Ding}
%\email{lei.ding@neel.cnrs.fr}
\affiliation{Universit\'e Grenoble Alpes, CNRS, Institut N\'eel, 38042 Grenoble, France}

\author{Claire V. Colin}
\author{Virginie Simonet}
\affiliation{Universit\'e Grenoble Alpes, CNRS, Institut N\'eel, 38042 Grenoble, France}

\author{Chris Stock}
\affiliation{School of Physics and Astronomy and Centre for Science at Extreme Conditions, University of Edinburgh, Edinburgh EH9 3FD, United Kingdgom}

\author{Jean-Blaise Brubach}
\author{Marine Verseils}
\author{Pascale Roy}
\affiliation{Ligne AILES-Synchrotron SOLEIL, 91190 Gif-sur-Yvette CEDEX, France}

\author{Victoria Garcia Sakai}
\affiliation{ISIS Pulsed Neutron and Muon Source, Rutherford Appleton Laboratory, Chilton, Didcot OX11 0QX, United Kingdom}

\author{Michael M. Koza}
\author{Andrea Piovano}
\author{Alexandre Ivanov}
\affiliation{Institut Laue-Langevin, 71 avenue des Martyrs, 38042 Grenoble, France}

\author{Jose A. Rodriguez-Rivera}
\affiliation{NIST Center for Neutron Research, National Institute of Standards and Technology, 100 Bureau Drive, Gaithersburg, Maryland 20899, USA}
\affiliation{Department of Materials Science, University of Maryland, College Park, Maryland, 20742, USA}

\author{Sophie de Brion}
\affiliation{Universit\'e Grenoble Alpes, CNRS, Institut N\'eel, 38042 Grenoble, France}

\author{Manila Songvilay}
\affiliation{Universit\'e Grenoble Alpes, CNRS, Institut N\'eel, 38042 Grenoble, France}
\affiliation{School of Physics and Astronomy and Centre for Science at Extreme Conditions, University of Edinburgh, Edinburgh EH9 3FD, United Kingdgom}

\date{\today}

\begin{abstract}
In metal-organic-framework (MOF) perovskites, both magnetic and ferroelectric orderings can be readily realized by compounding spin and charge degrees of freedom. The hydrogen bonds that bridge the magnetic framework and organic molecules have long been thought of as a key in generating multiferroic properties. However, the underlying physical mechanisms remain unclear. Here, we combine neutron diffraction, quasielastic and inelastic neutron scattering, and THz spectroscopy techniques to thoroughly investigate the dynamical properties of the multiferroic MOF candidate [CH$_3$NH$_3$][Co(HCOO)$_3$] through its multiple phase transitions. The wide range of energy resolutions reachable by these techniques enables us to scrutinize the coupling between the molecules and the framework throughout the phase transitions and interrogate a possible magnetoelectric coupling. Our results also reveal a structural change around 220 K which may be associated with the activation of a nodding donkey mode of the methylammonium molecule due to the ordering of the CH$_3$ groups. Upon the occurrence of the modulated phase transition around 130 K, the methylammonium molecules undergo a freezing of its reorientational motions which is concomitant with a change of the lattice parameters and anomalies of collective lattice vibrations. No significant change has been however observed in the lattice dynamics around the magnetic ordering, which therefore indicates the absence of a substantial magneto-electric coupling in zero-field.

%The spin excitations were successfully modeled using a set of exchange interactions and a large in-plane anisotropy. 
%We propose that spin and lattice degrees of freedom might be weakly coupled as indicated by the anomalous change of the intensity of a phonon mode of methylammonium around the magnetic ordering temperature. 

\end{abstract}

\maketitle

\section{Introduction} 

\begin{figure*}
\centering
\includegraphics[scale=0.5]{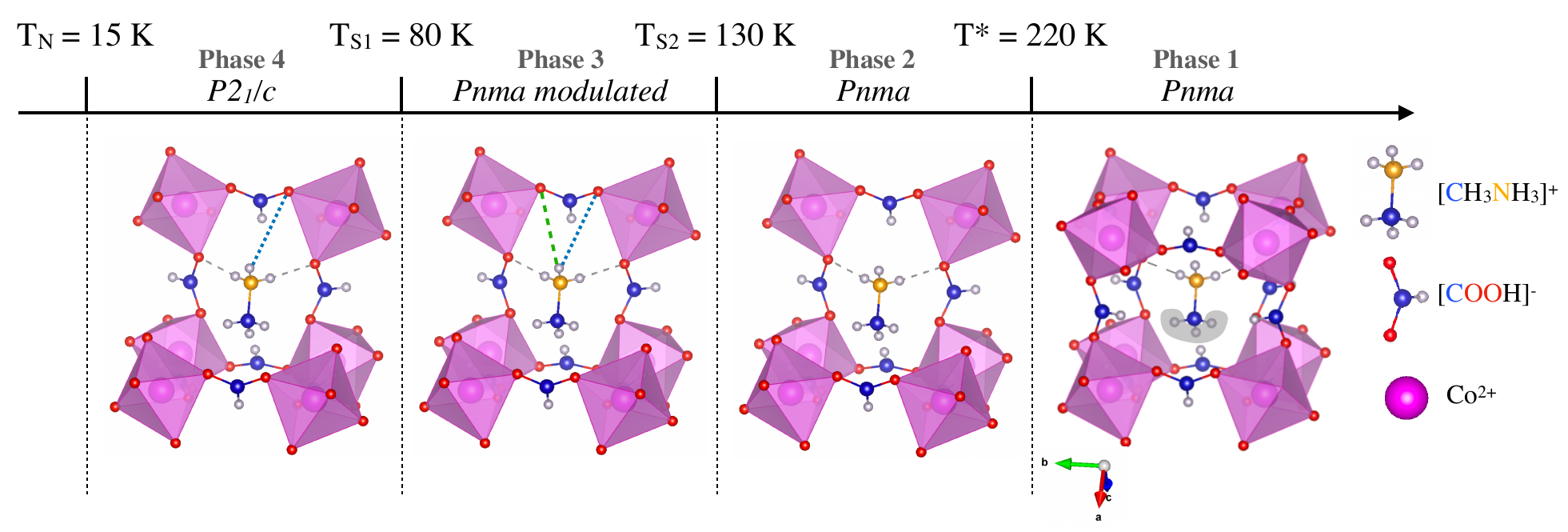}
\caption{Summary of the structural distortions of [CH$_3$NH$_3$][Co(HCOO)$_3$] as a function of temperature. The dashed lines represent the different hydrogen bonds linking the nitrogen atom of the methylammonium molecules [CH$_3$NH$_3$]$^+$ to the CoO$_6$ octahedra (for more clarity, two CoO$_6$ octahedra were removed in the 3 left panels). The different colors (grey, green, blue) represent the different lengths of these bonds, two of which vary as a function of temperature. The grey shaded area around the carbon atom in the righ panel illustrates the site disorder described in the text.}\label{fig:summary}
\end{figure*}

Metal-organic-framework (MOF) compounds represent extremely versatile materials in which organic and inorganic structural components may be hybridized, engendering a wide variety of functional properties. 
Indeed, MOF perovskites have recently received a lot of attention as they exhibit fascinating properties such as photovoltaic and optoelectronic applications \cite{Stranks2015, Frost2016}, gas storage capabilities \cite{Furukawa2013} and multiferroic properties \cite{Saparov2016, Li2017, Ma2020}. Deriving from the typical perovskite  structure ABX$_3$, the hybrid organic-inorganic compounds comprise a framework of BX$_6$ octahedra, where $B$ can be a metal or a transition metal and $X$ a halide or an organic linker, such as formate (COOH$^{-}$). This framework hosts a molecular cation on the $A$-site, such as methylammonium (CH$_{3}$NH$_{3}^{+}$, noted MA), interacting with the framework through hydrogen bonds, as illustrated in Fig. \ref{fig:summary}. 

In particular, the role of the hydrogen bonds and the interplay between the A-site molecules and the framework in their exceptional functional properties have been largely discussed \citep{Di2013,Even2014,Frost2016,Quarti2014}. This has been extensively studied in lead-halide MOF perovskites in order to rationalize the role of the order-disorder transition of the MA molecule in the photovoltaic properties of these materials \cite{Swainson2015,Leguy2015,Brown2017,Songvilay2019}. On the other hand, it has been shown recently that inorganic lead halide perovskites feature similar efficiencies and therefore, the importance of hydrogen bonding and the molecular nature of the A-site are still debated \cite{Kulbak2015,Songvilay2019Cs}. While hydrogen bonding may not be essential in lead halide perovskites, it has been shown that it is about three times stronger in formate perovskites \cite{Svane2017}, which interrogates the strength of the coupling between the A-site cation and the framework in other types of MOF perovskites. 

%However, the physics of molecules in the enhanced photovotaic effect in these MOFs remains uncertain probably due to the lie in the fact that the hydrogen bonding is weak in lead-halide perovskites, and a structure-property relationship is hard to establish. Since the hydrogen bonding  than in halide perovskites , the molecular-framework interaction in the former should be stronger.

This is what motivates the search for new multiferroic materials among the formate-based hybrid perovskite compounds. In these materials, on the one hand, the asymmetric molecules on the A-site (methylammonium, dimethylammonium or formamidinium) carry a dipolar moment. As the molecules undergo an order-disorder transition at low temperature, a ferroelectric or antiferroelectric arrangement depending on the orientations of the molecular groups becomes feasible. On the other hand, a magnetic transition metal ion on the $B$-site may give rise to a long-range magnetic order which may couple to the ferro/antiferroelectric order. Indeed, in a number of magnetic formate perovskites, the electric dipole and spin ordering have been found experimentally, but at different temperatures \cite{Jain2009,Fu2011,Tian2016,Xu2011,Miroslaw2017,Hughey2017,Hughey2018,Ma2020}. However, the mechanism of the possible magneto-electric coupling in these formate perovskites is unclear, despite the effort that has been made to investigate the coupling between the spin and charge degrees of freedom in the magnetically ordered phase \cite{Hughey2017}. 

More recently, the MOF perovskite [CH$_3$NH$_3$][Co(HCOO)$_3$] (MA-Co) has been suggested to host a magnetic-ordering induced multiferroic behavior \cite{Claudia2016,Pato2016}. Below 4 K, it has been shown that a small change in the electric polarization ($\Delta P = 0.28 \mu C/m^2$) appears when applying a magnetic field of 4 T parallel to the [1 0 1] direction, concomittant with a change in magnetization. A subsequent work found that it also undergoes a paraelectric to antiferrroelectric transition around 78 K, which coincides with an orthorhombic to monoclinic distortion, followed by a canted antiferromagnetic ordering below 15.9 K \cite{Mazzuca2018}. In the antiferroelectric phase, the MA groups orient in an opposite manner, giving rise to zero net electric polarisation. A comprehensive neutron diffraction work unveiled an additional intermediate structural distortion with a modulated crystal structure between 128 K and 78 K in MA-Co \cite{Canadillas2019}, as summarised in Fig. \ref{fig:summary}. 
%Although the static structures of MA-Co in these phases have been determined, the nature of the multiple phase transitions remains unknown. 
In order to investigate the mechanisms involved in these phase transitions, we employ neutron diffraction, quasielastic and inelastic neutron scattering, and THz spectroscopy to study the molecular dynamics, lattice dynamics and spin excitations spanning a wide range of energy scales. These experimental techniques allow us to study the role of molecules on the crystal structural and magnetic phase transitions and to quest for a possible magnetoelectric coupling in MA-Co.

\section{Methods}
Polycrystalline samples and single crystals were synthezised using the method described in \cite{Mazzuca2018}. Note that the samples were fully protonated (no deuteration).
Powder neutron diffraction experiments were performed on the two-axis diffractometer D1B at Institut Laue Langevin (ILL, France), from 1.5 K to 300 K using the neutron wavelength 2.52 \AA. Neutron data were analyzed by a profile matching analysis using the FULLPROF suite program \cite{Fullprof}.  

Quasielastic neutron scattering measurements were carried out on the indirect-geometry time-of-flight spectrometer OSIRIS at ISIS facility (UK). Graphite PG002 analyzers were used to produce a fixed final energy E$_{f}$~=~1.84 meV. With an incident white beam of neutrons on the sample, the default dynamic range of the instrument is $\pm$ 0.5 meV. Quasielastic neutron data were collected in the temperature range of 5-290 K using a powder sample. The time-of-flight neutron spectrum was corrected for detector efficiency and then normalized against the monitored beam flux at the sample space.

Powder inelastic neutron scattering experiment was done on the thermal-neutron direct-geometry time-of-flight (TOF) spectrometer PANTHER at the ILL \cite{DOIPanther}. TOF neutron scattering data were measured in the temperature range of 1.5-300 K with different incident energies E$_i$ = 19, 40 and 76 meV. Complementary measurements were carried out on a large single crystal in the (H~0~L) scattering plane, using the thermal triple-axis spectrometer IN8 with a constant energy E$_{f}$~=~14.7~meV \cite{DOIIN8}.

Magnetic excitations were investigated using inelastic neutron scattering on two single crystals oriented in the (H~0~L) and the (H~K~H) scattering planes respectively, using the cold-triple axis spectrometer MACS located at NIST (United States). For both orientations, measurements were carried out with a fixed E$_{f}$~=~3.7~meV and a BeO filter on the scattering side to reduce the background.

High resolution THz spectroscopy measurements were performed on the AILES beamline at SOLEIL Synchrotron (France) in the energy range 10 - 100 cm$^{-1}$ (1.25 - 12.5 meV) where the THz wave is naturally polarized. Plaquettes of millimeter size and with a thickness of about 500 $\mu$m were oriented with the $b$-axis lying in the plane of the plaquettes, in order to probe the (e $\parallel$ b, h $\perp$ b) and (e $\perp$ b, h $\parallel$ b) geometries of the THz electric (e) and magnetic (h) fields. THz absorption spectra at different temperatures from 5 K to 300 K were collected in the transmission configuration with a resolution of 0.5 cm$^{-1}$.

\section{Results}

\subsection{Temperature dependence of the phase transitions}

\begin{figure}
\includegraphics[scale=0.37]{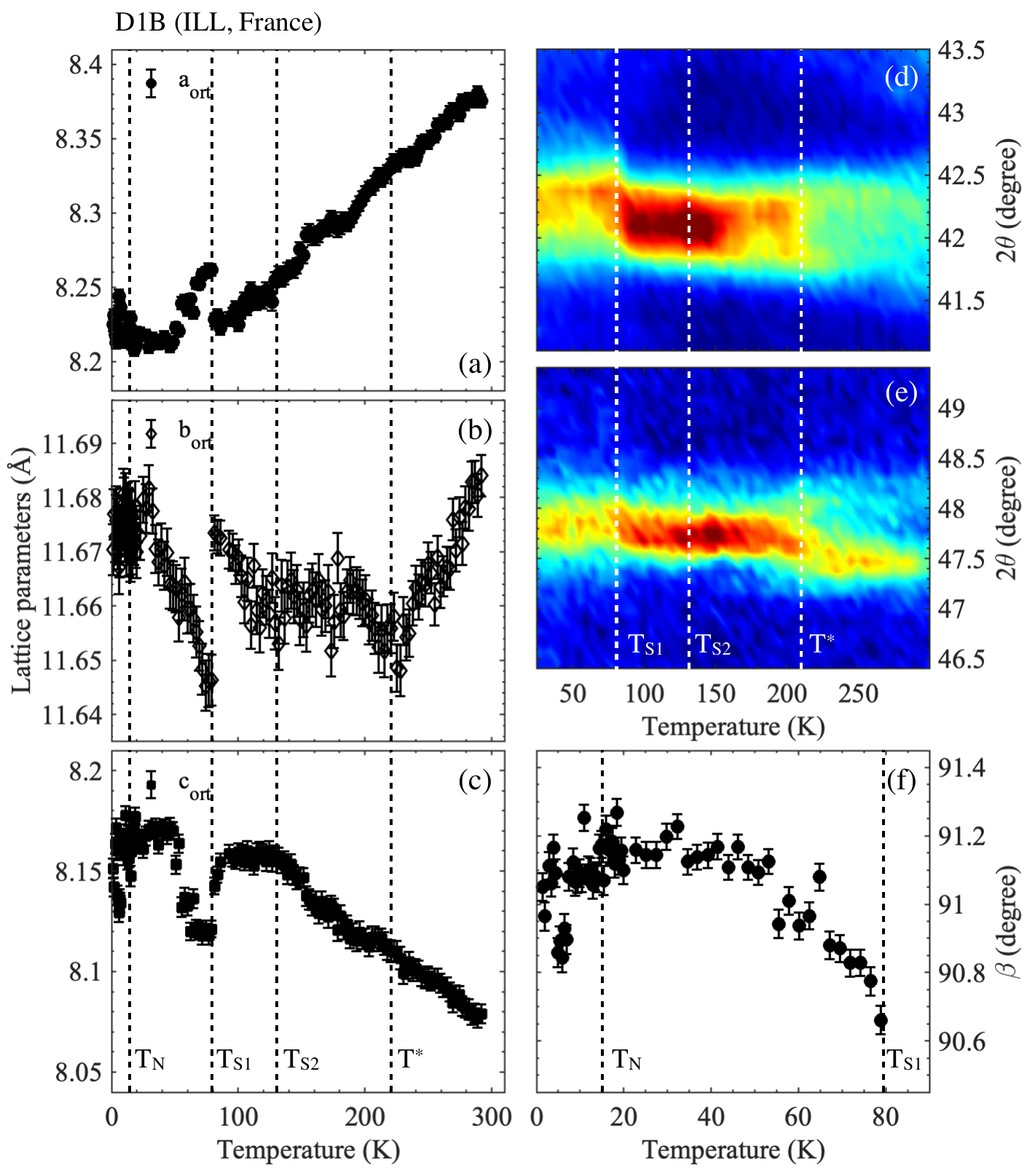}
\caption{(a-c) Temperature dependence of the lattice parameters from the powder neutron diffraction, refined in the orthorhombic space group. (d-e) Contour plot versus temperature of the powder neutron intensities of two Bragg reflections. (f) Temperature dependence of $\beta$ monoclinic angle of the low temperature phase. T$_{S1}$, T$_{S2}$ and T* mark the three structural phase transitions and changes while T$_N$ indicates the magnetic transition.}\label{fig:d1b}
\end{figure}

In order to investigate the phase transitions in [CH$_3$NH$_3$][Co[HCOO]$_3$], powder neutron diffraction measurements were carried out in the temperature range of 1.5-300 K.
The neutron diffraction contour plots of selected Bragg reflections shown in Figures \ref{fig:d1b} (d-e) disclose a variation of their intensity and 2$\theta$ angle signaling two phase transitions: one previously reported transition between monoclinic and orthorhombic symmetries at T$_{S1}$~=~80~K \citep{Canadillas2019} and an additional distortion at higher temperature T$^{*}$~=~220~K. 
In order to characterize them, the lattice parameters were fitted in a profile matching refinement as a function of temperature using the previously reported structural models for the orthorhombic (P$nma$) and monoclinic (P$2_1/c$) phases with a lattice transformation a$_{ort}$ = -b$_{mono}$, b$_{ort}$ = -c$_{mono}$ and c$_{ort}$ = a$_{mono}$ \cite{Mazzuca2018,Canadillas2019}. The temperature-dependent lattice parameters, displayed in Fig. \ref{fig:d1b} (a-c), show two structural anomalies at T$_{S2}$ = 130 K, corresponding to the transition between the high temperature orthorhombic $Pnma$ phase to an orthorhombic modulated $Pnma$ phase, and at  T$_{S1}$ = 80 K, corresponding to the transition to the monoclinic phase \citep{Canadillas2019}. Indeed, these transitions are marked by the changes in the lattice parameters $a_{ort}$, $b_{ort}$, $c_{ort}$ and the increase of the value of the $\beta$ angle of the monoclinic phase around T$_{S1}$ (Fig.\ref{fig:d1b} f).  At higher temperature, and as reflected in Fig. \ref{fig:d1b} (b), (d) and (e), a clear change of slope in the temperature dependence of the $b$ parameter is observed at T$^{*}$ = 220 K, as well as a splitting in some nuclear Bragg reflexions, illustrated in the colormaps (Fig. \ref{fig:d1b} (d) and (e)).
While the phase transitions at T$_{S1}$ and T$_{S2}$ have been investigated in the previous works, the structural modification at T$^*$ has not been reported yet and will be discussed in the following sections.

\subsection{Reorientational dynamics of MA molecules}

\begin{figure}
\includegraphics[scale=0.34]{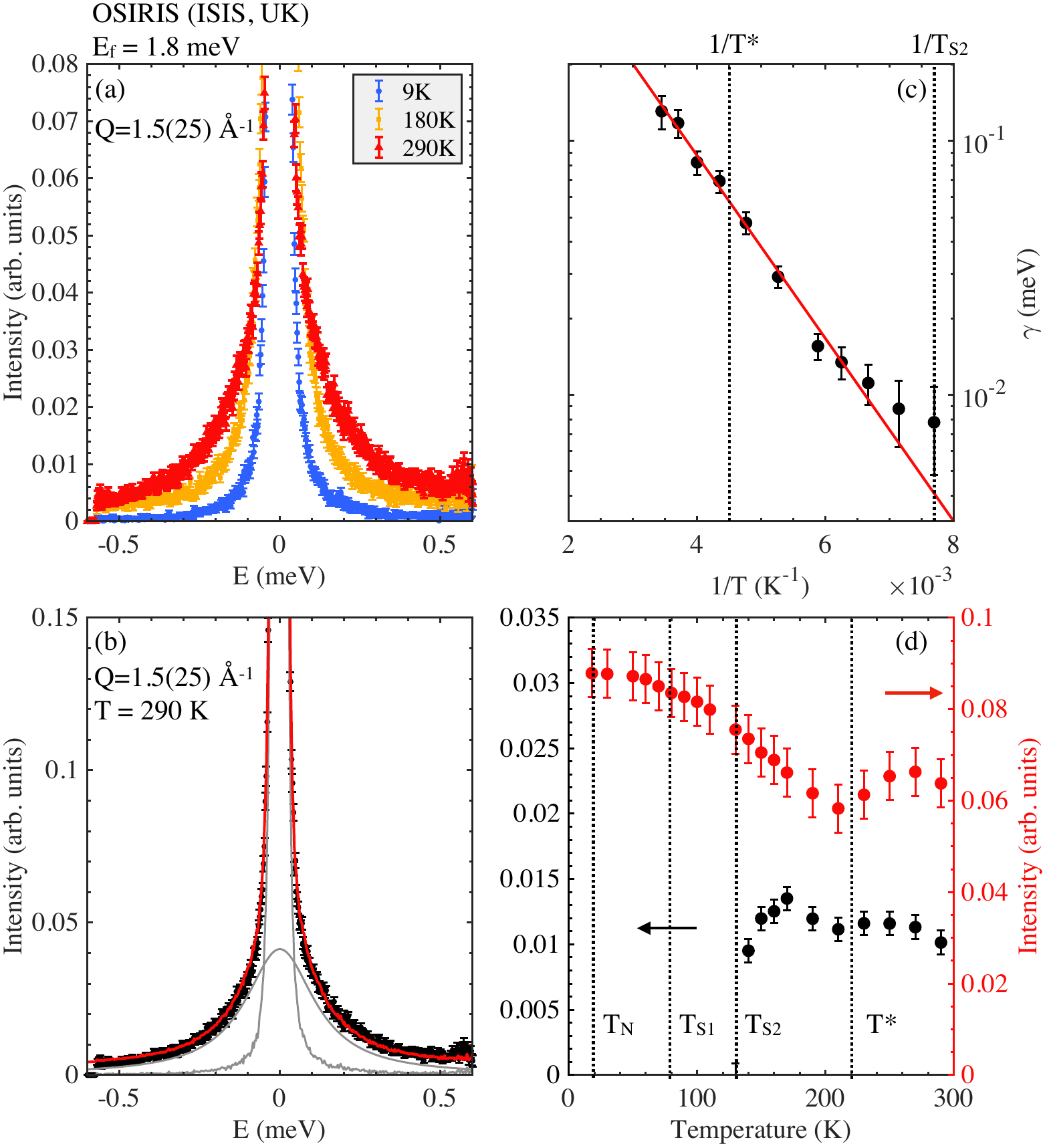}
\caption{(a) Quasielastic incoherent neutron scattering spectra measured at 9, 180 and 290 K. (b) Quasielastic incoherent neutron scattering spectrum at 290 K showing the fit composed a quasi-elastic dynamic component and an elastic resolution-limited component (in grey). The total fit to the model described in the text is given by the red curve. (c) The Arrhenius plot of the quasielastic energy linewidth as a function of temperature. (d) Temperature dependence of the elastic intensity (red circles) and quasi-elastic intensity (black circles).}\label{fig:qens}
\end{figure}

To further characterise the possible coupling between the MA molecules and the framework, the dynamics of MA molecules was studied using high resolution quasielastic incoherent neutron scattering on a powder sample of MA-Co, using the OSIRIS backscattering spectrometer. The technique is well suited to study the reorientational motions or jumps of MA molecules via the large incoherent cross-section of the hydrogen atoms in the methylamonium molecule, compared to the other atoms in MA-Co. The motions of hydrogen atoms of formate molecules may also slightly contribute to the incoherent neutron scattering; however, since each formate molecule connects two CoO$_6$ octahedra and the framework is well ordered, as shown in Fig. \ref{fig:summary}, its contribution appears to be minor and neglected in the energy dependence analysis \citep{Lu2021}. 

We first examine the low-energy spectra as a function of temperature, in order to determine the proper model to describe the observed quasielastic scattering data. The neutron spectra, measured at 9 K, 180 K and 290 K, were obtained by integrating the scattered intensity over a $Q$ range of 1.25-1.75~\AA~and are shown in Fig. \ref{fig:qens} (a). The spectrum at 9 K is well described by a single Gaussian function whose width corresponds to the instrumental resolution ($\sim$ 0.02 meV). No additional scattering was observed, which means that no dynamics are visible at this temperature within the timescale accessible on OSIRIS. With increasing temperature, the profile of the neutron spectrum broadens and reveals an energy profile with two components. At 290 K, it consists of an elastic peak superimposed on a weaker and broader peak. The former component's energy width is resolution limited, just like the data at 9 K, while the latter extends to at least $\pm$ 0.5 meV, suggesting that the neutron spectra in MA-Co is characterized by a static and a dynamic timescale. The spectra were then fitted to a model containing both components in energy, represented respectively by one delta-function and one Lorentzian \citep{Line1994,Songvilay2019}:

\begin{eqnarray}
I(Q,E) = I_{el}(Q)\delta(E) + \dfrac{I_{dyn}(Q)}{1+(E/\gamma(T))^2}
\end{eqnarray}
the whole $I(Q, E)$ function being convoluted with the experimentally determined resolution function given by the spectrum measured at 9 K.
I$_{el}$ and I$_{dyn}$ are the static and dynamic amplitudes, respectively, $\gamma$ is the half-width at half-maximum in energy of the dynamic component, inversely proportional to the lifetime of the molecular fluctuations $\tau$ $\sim$ 1/$\gamma$. As a representative, the fitted results at 290 K are shown in Fig. \ref{fig:qens} (b). At given temperatures, the energy linewidth $\gamma$ is independent of the momentum transfer $Q$ (not shown), indicating that there is no measurable diffusion of the molecules in the sample \cite{Songvilay2019}.  
Figure \ref{fig:qens} (d) shows the fitted elastic and quasielastic intensity in the temperature range of 15-300 K. We found that below T$_{S2}$ all neutron spectra are accounted for by only the elastic component, signifying that all molecular motions are practically frozen within the timescale of the experiment. On the other hand, above T$_{S2}$, the quasielastic dynamical component associated with the molecular dynamics shows a rapid increase before reaching saturation around T = 180 K. Above T$^{*}$, an additional flat background intensity appears (Fig. \ref{fig:qens} a), which indicates the onset of a dynamical component whose timescale falls outside of the dynamic range of the instrument.
The quasielastic linewidth $\gamma$ as a function of temperature shows a monotonic increase above T$_{S2}$ with increasing temperature. We observe no anomaly in the linewidth around T$^{*}$. The temperature-dependent quasielastic linewidth was fitted to an Arrhenius law $\gamma$ (T) = $1/\tau_0$exp(-E$_a$/k$_B$T), as shown in Fig. \ref{fig:qens} (c). The fit was performed in the high temperature regime, as the linewidth slightly deviates from this law around T$_{S2}$. A characteristic time  $\tau_0$ = 17 ps  and an activation energy E$_a$~=~830(30)~K was extracted from the fit.  While the activation energy is larger than those found in metal-inorganic lead-halide perovskites (ranging from 200 to 600 K depending of the halide \cite{Songvilay2019, Brown2017,Leguy2015, Chen2015}), where the cavities of the framework are also filled by the MA molecules, the characteristic time of the molecular fluctuations is of the same order of magnitude. \citep{Swainson2015, Songvilay2019}. 
To summarize, in this energy range, the lattice dynamics is dominated by the onset of a quasi-elastic component that appears above T$_{S2}$, which is associated with the dynamics of the methylammonium molecule. As illustrated in Fig.~\ref{fig:summary} and according to ref. \cite{Canadillas2019}, the molecule is bound to the organic framework via two out of three hydrogen bonds surrounding the nitrogen atom, while the methyl group [CH$_{3}$] remains free. Although an additional lattice distortion has been detected at T$^{*}$ in the neutron diffraction study, no particular anomaly is observed in the quasi-elastic intensity or linewidth at this temperature.

%the quasielastic neutron scattering results show the onset of a dynamical component which intensity and linewidth increase monotonically as a function of temperature, with no particular change around T$^{*}$. Above T$^{*}$ however, an additional dynamical component appears, which timescale lies outside the instrument resolution. 

%The large activation energy in MA-Co {\color{blue}may be} attributed to {\color{blue}a} stronger coupling between the molecules and the framework as a consequence of its stronger hydrogen bonding \cite{Svane2017}. This activation energy appears to be associated with the potential barrier for relative rotations of CH$_3$ \textcolor{red}{group, which is unbound to the framework cage against the NH$_3$ group of the same MA molecule that is attached to the framework by hydrogen bondings. (Est-ce vrai? References?. Ce serait bien d'avoir ici un petit résumé des résultats de dynamique moléculaire pour les différentes transitions. Pas très clair ce qui se passe à T* et TS2.)}.

\subsection{Low-energy lattice dynamics and magnetic excitations}

\begin{figure*}
\centering
\includegraphics[scale=0.45]{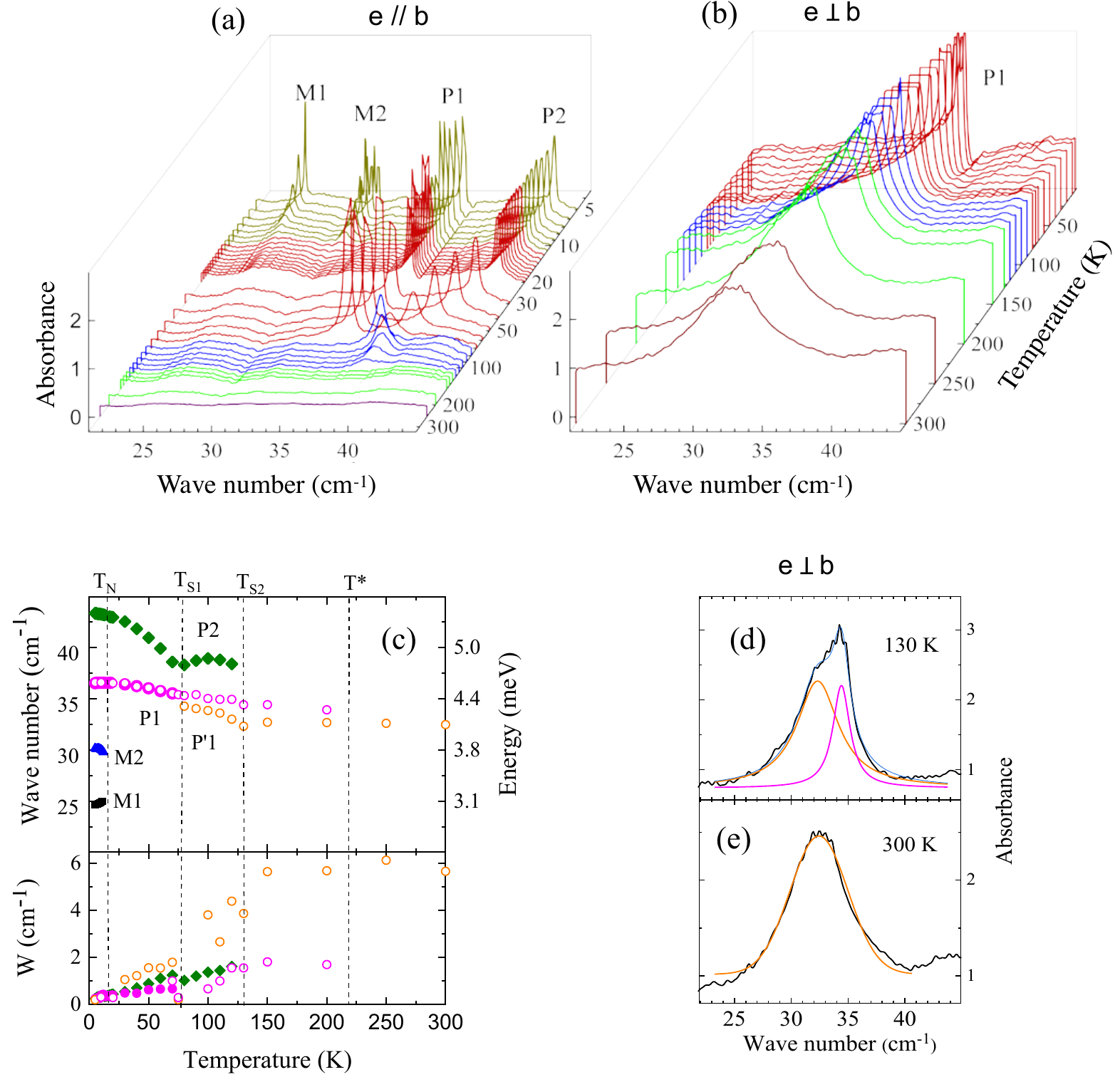}
\caption{THz spectroscopy of [CH$_3$NH$_3$][Co(HCOO)$_3$]. Absorbance spectra recorded at various temperatures with two different polarization for the incident beam : e$//$b$_{ort}$ (a) and e $\perp$ b$_{ort}$ (b). Note that the absorbance is saturated above a maximum value of 3. (c) Temperature dependence of the peak positions (upper panel) and width (lower panel) for e$//$b$_{ort}$ (closed symbols) and e $\perp$ b$_{ort}$ (open symbols) in cm$^{-1}$ and in meV.  (d-e) Fits of the absorbance spectrum measured at 130 and 300 K for e $\perp$ b$_{ort}$. }\label{fig:THz}
\end{figure*} 

THz spectroscopy is capable of probing the characteristic energies for electric lattice dynamics, i.e. infrared active phonons thanks to the electric component (e) of the electromagnetic wave. In both the orthorhombic $Pnma$ and monoclinic $P2_1/c$ phases, 249 optical modes are expected, that are all singlets. It is their infrared activity that differ so that new modes are expected to appear at the phase transition, mostly associtated to the heavy octahedra around Co$^{2+}$ ions. Additionally, THz spectroscopy can also probe spin dynamics (such as magnons) thanks the magnetic (h) components of the polarized electromagnetic wave keeping in mind that e $\perp$ h. Figure \ref{fig:THz} shows the THz spectra as a function of temperature measured using two configurations e~$\parallel$~b$_{ort}$ (h~$\perp$~b$_{ort}$) and e~$\perp$~b$_{ort}$ (h~$\parallel$~b$_{ort}$). At 5 K, four excitations at 24.6 cm$^{-1}$ (M1), 29.8 cm$^{-1}$(M2), 36.6 cm$^{-1}$ (P1) and 43.3 cm$^{-1}$(P2) are clearly seen as sharp peaks in the spectra. The excitations M1 and M2 that are excited when e $\parallel$ b$_{ort}$ and h $\perp$ b$_{ort}$, are of magnetic origin since they disappear above T$_N$. They also agree very well with the magnetic excitations measured by neutron scattering (discussed in the following section) at an equivalent zone center of the Brillouin zone. They are associated with spin fluctuations perpendicular to b$_{ort}$. The two other excitations P1 and P2 are of phononic origin and are noticeably affected by T$_{S1}$, T$_{S2}$ and T$^*$. The P2 mode has the simplest behavior as it is only present in the channel  e~$\parallel$~b$_{ort}$. With increasing temperature, it softens substantially and shows a minimum around T$_{S1}$, before disappearing above T$_{S2}$ with a noticeable broadening. The P1 mode is activated in both channel e~$\parallel$~b$_{ort}$ and e~$\perp$~b$_{ort}$ up to T$_{S1}$ and softens slightly. Above T$_{S1}$, it remains only when e~$\perp$~b$_{ort}$ with a double structure: another peak P'1 emerges at slightly lower energy than P1 with a substantial broadening  with increasing temperature (Fig. \ref{fig:THz} d). These two modes display different behaviors, as P'1 is still present above T$^*$ while P1 gradually disappears up to  T$^*$ (Fig. \ref{fig:THz} c). These new modes, that are clear signatures of the different structural phases, are certainly related to the heavy octahedra of the structure, although assigning these modes to specific vibrational modes is not  straight forward: in this low energy range, the calculations performed on the lead halides are very sensitive to the space group \cite{Leguy2015,Songvilay2019}. Note that peak broadening occurs when disorder is present and/or the excitation life time is reduced. The variation of  energy of the different modes as a function of temperature is summarized in Fig. \ref{fig:THz} (c). All the phases can be clearly distinguished and are enlightened in different colors in Fig. \ref{fig:THz} (a) and (b). 

In the THz range, we can then identify signatures of the different phases. The magnetic order is characterized by magnons M1 and M2 while the monoclinic phase below T$_{S1}$ is characterized by phonons P1 and P2. In the intermediate  orthorhombic modulated $Pnma$ phase that is present between T$_{S1}$ and T$_{S2}$, another broad phonon P'1 appears close to P1. The transition to the high temperature orthorhombic $Pnma$ phase shows up with the disappearance of  P2. Finally, above T$^*$, P1 disappears and only P'1 remains.  Note that magnons and phonons are well separated in energy, which suggests that, spin and lattice degrees of freedom are not coupled, as indicated by the lack of change in the phonon modes around T$_N$.

%With increasing temperature, mode P1 softens slightly and disappears around T$_{S1}$ for e$\parallel$ b$_{ort}$. It is present similarly for  e $\perp$ b$_{ort}$ belowT$_{S1}$, but remains above when a double structure prevails withthe appearance of a broader peak that is the only one to remain above T$^*$ . \textcolor{red}{(pourquoi les largeurs des deux pics sont tres differentes?)}. Simultaneously, mode P2 softens substantially and shows a minimum around T$_{S1}$. It finally disappears above T$_{S2}$. The variation of the \textcolor{red}{energy of the} different modes as a function of temperature is summarized in Fig. \ref{fig:3}c. All the phases can be clearly distinguishable and are enlightened} in different colors in Fig. \ref{fig:3}a and b. Note that magnons and phonons are well separated in energy, suggesting that, if spin and lattice degrees of freedom are coupled, it is not through the infrared active low energy phonon modes. \textcolor{red}{(il faut une correspondance meV  cm-1 sur les figures sinon, on ne peut pas se repérer). Quelle conclusion sur cette partie?}

\subsection{High-energy lattice dynamics}

\begin{figure*}
\centering
\includegraphics[scale=0.45]{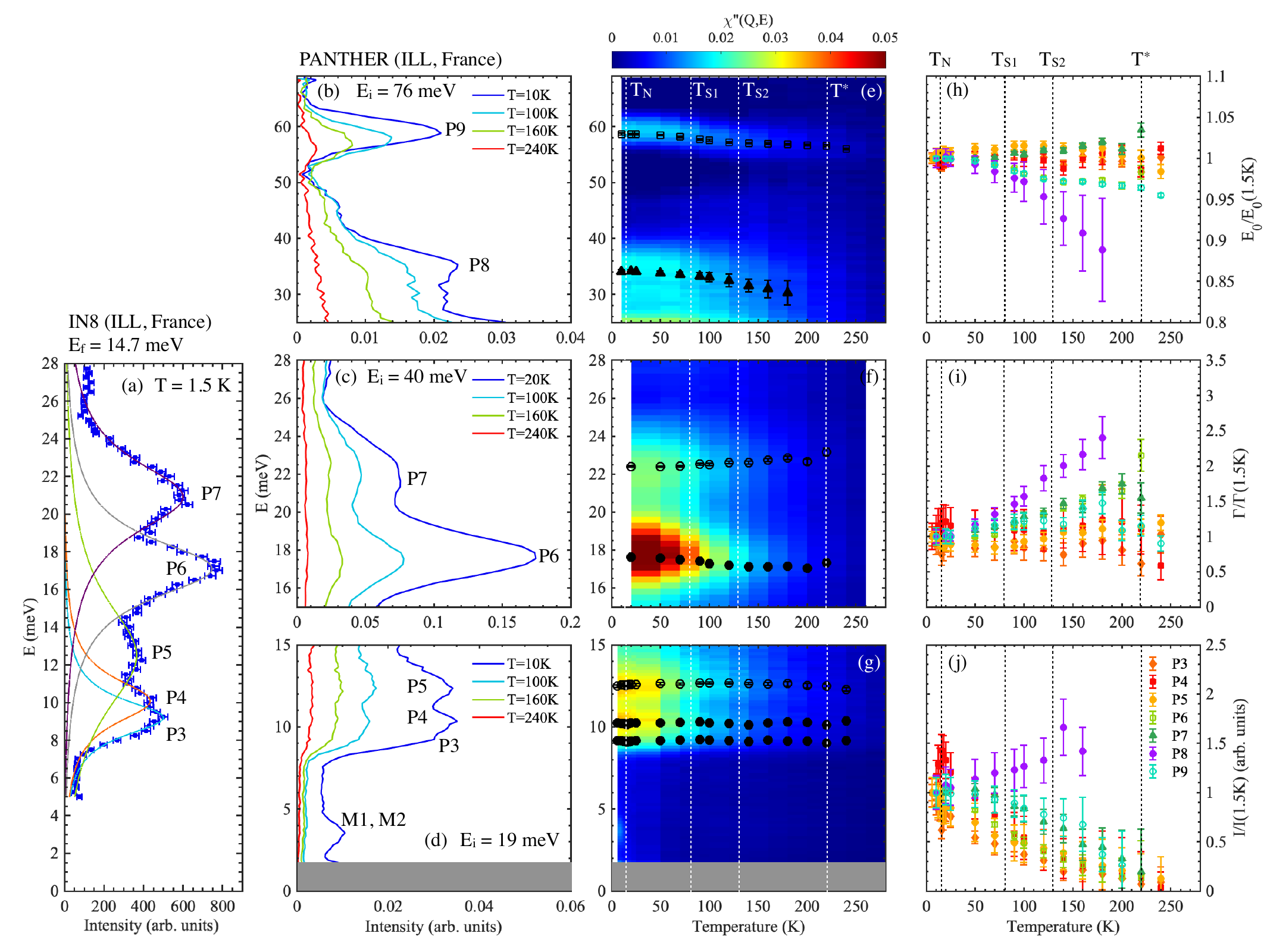}
\caption{(a) Constant-Q scan (Q = (4~0~1)) measured on IN8 at T = 1.5 K. The lines illustrate the fit of each phonon mode to a harmonic oscillator, as described in the text. (b-d) Neutron inelastic spectra integrated over the entire Q range at several temperatures. (e-g) Temperature dependence of the Q-integrated spectra. The data points represent the characteristic energy of each mode, fitted with the model described in the text. The incident energies are E$_i$ =19 (d,g), 40 (c,f) and 75 meV (b,e). (h-j) Relative temperature variation with respect to the T = 1.5 K reference of the energy position (h),  the linewidth (i) and the intensities (j) of the phonon modes.}\label{fig:panther}
\end{figure*}

Higher energy phonon modes associated with the molecular and framework lattice vibrations were also measured using inelastic neutron scattering on the thermal triple-axis spectrometer IN8 (Fig.\ref{fig:panther} a) and the time-of-flight neutron spectrometer PANTHER with different incident energies E$_i$ = 19, 40 and 75 meV, allowing to access different dynamic ranges (Fig.\ref{fig:panther} b-g).
Figures~\ref{fig:panther} (a-c) and (d-f) present the Q-integrated inelastic neutron intensity measured as a function of temperature, between 1.5 and 300 K, with incident energies E$_i$~=~19, 40 and 75 meV, from bottom to top. The low temperature data show four modes (P3, P4, P5 and P6) below 20 meV, 2 modes (P7 and P8) between 20 and 40 meV and one mode around 60 meV (P9). Interestingly, these modes are comparable to the ones previously observed through neutron, Raman and infrared spectroscopy in the non-magnetic lead halide perovskites MAPbX$_{3}$ (X = I, Br, Cl) \cite{Leguy2016, Songvilay2019, Swainson2015, Quarti2014, Glaser2015, Brivio2015}, and which were confirmed by DFT calculations. Referring to the previous works, we may thus tentatively assign the different modes measured on PANTHER in MA-Co to specific octahedral and molecular vibrations. When comparing to the non-magnetic lead halide perovskites, the first noticeable difference is the apparent absence of phonon modes below 8 meV which were assigned to PbX$_{6}$ twisting and distortions of the inorganic framework in the lead-halide perovskites. Our THz spectroscopy measurements show that these modes are indeed present around 5 meV: these are excitations P1 and P2 around 40 cm$^{-1}$ (see Fig. \ref{fig:THz} and discussion in the previous section), but they are too weak to be detected in our neutron measurements. The observed higher energy modes, in agreement with the above-mentioned previous spectroscopy studies, involve the MA molecules: the modes P3, P4 and P8, observed around 8, 10 meV and 35 meV, are referred to as ``nodding donkey'' modes, which correspond to rotational vibrations of the molecule around the nitrogen atom for the two lowest modes (P3 and P4) and around the carbon atom for the highest one (P8) \cite{Leguy2016}. These modes and molecular motions have been discussed in \cite{Swainson2015, Brown2017, Quarti2014, Brivio2015} as being directly affected by the inorganic framework distortions in the lead-halide family since they involve the hydrogen bonds between the MA cations and the CoO$_{6}$ octahedra. Hence, these phonon modes may provide information on the coupling between the MA molecules and the framework. The two modes P6 and P7 between 15 and 30 meV correspond to lurching of the MA molecule, and the phonon mode P9 around 60 meV involves internal motions of the molecule as it is associated with torsions around the C-N axis.
In order to get further insight into the role of the different molecular motions in the structural transitions, it is interesting to examine the phonon temperature dependence, which is displayed in figure \ref{fig:panther}. To extract the energy position, the width and the intensity of each phonon mode as a function of temperature, the inelastic spectra were corrected from the Bose factor and a constant background, and the phonon modes were fitted to a harmonic oscillator  model (as illustrated in Fig. \ref{fig:panther} a): 
\[\chi''(\vec{Q},\omega) = I \left(\frac{\Gamma}{\Gamma^2 + (\omega-E_0)^2} - \frac{\Gamma}{\Gamma^2 + (\omega + E_0)^2}\right)\]
where $I$ is a constant, $\Gamma$ is the phonon linewidth and $E_{0}$ is the energy position.
With the fitted parameters, we obtained the relative energy position, linewidth and intensity of the phonon modes as compared to the parameters fitted at 1.5 K, which are plotted in Fig. \ref{fig:panther} (g-i). 

Although multiple phase transitions occur, no particular anomaly is observed in the inelastic data as a function of temperature, except for the mode at 35 meV (P8). Indeed, as illustrated in Fig. \ref{fig:panther} (g-i), above T$_{S2}$, this mode seems to soften with increasing temperature and shows a clear broadening, while the other phonon modes remain unaffected up to T$^{*}$. Moreover, while the intensity of the other phonon modes decrease when heating, the intensity of the P8 mode increases before disappearing between T$_{S2}$ and T$^{*}$.
T$_{S2}$ corresponds to the modulated phase transition temperature which produces, upon cooling, a modulation in the bond length between the hydrogen atoms surrounding the nitrogen atom of the MA molecules and the CoO$_{6}$ octahedra (see figure \ref{fig:summary}) \cite{Canadillas2019}. The inelastic scattering data therefore show that this modulation of these hydrogen bonds significantly affects the motions of the nitrogen atoms around the carbon atom of the MA molecule.
By contrast, the modes observed at 8 and 10 meV (P3 and P4) do not show particular temperature dependence, as they involve the free motions of the carbon atom around the nitrogen atom of the molecule, as the hydrogen atoms of the [CH$_3$] group do not establish hydrogen bonds. 
This asymmetry in the molecular vibrations is particularly interesting as it provides further evidence for the modulated phase and reveals information on the coupling between the MA molecules and the framework. The appearance of the``nodding donkey'' motion around the carbon atoms below T$^*$ (P8 mode) may suggests that T$^*$ involves a structural modification that affects the carbon atom of the MA molecule. Our neutron inelastic data shows that the lower temperature structural transition from the orthorhombic to monoclinic symmetry at T$_{S1}$ and the magnetic transition at T$_{N}$ do not influence the high energy lattice dynamics.

%The fact that this specific molecular motion, directly involving the hydrogen bonds, is strongly affected by the modulated phase attests to a strong coupling between the methylammonium molecule and the host cobalt-formate framework. 
 
\subsection{Magnetic excitations and modeling}

\begin{figure*}
\centering
\includegraphics[scale=0.45]{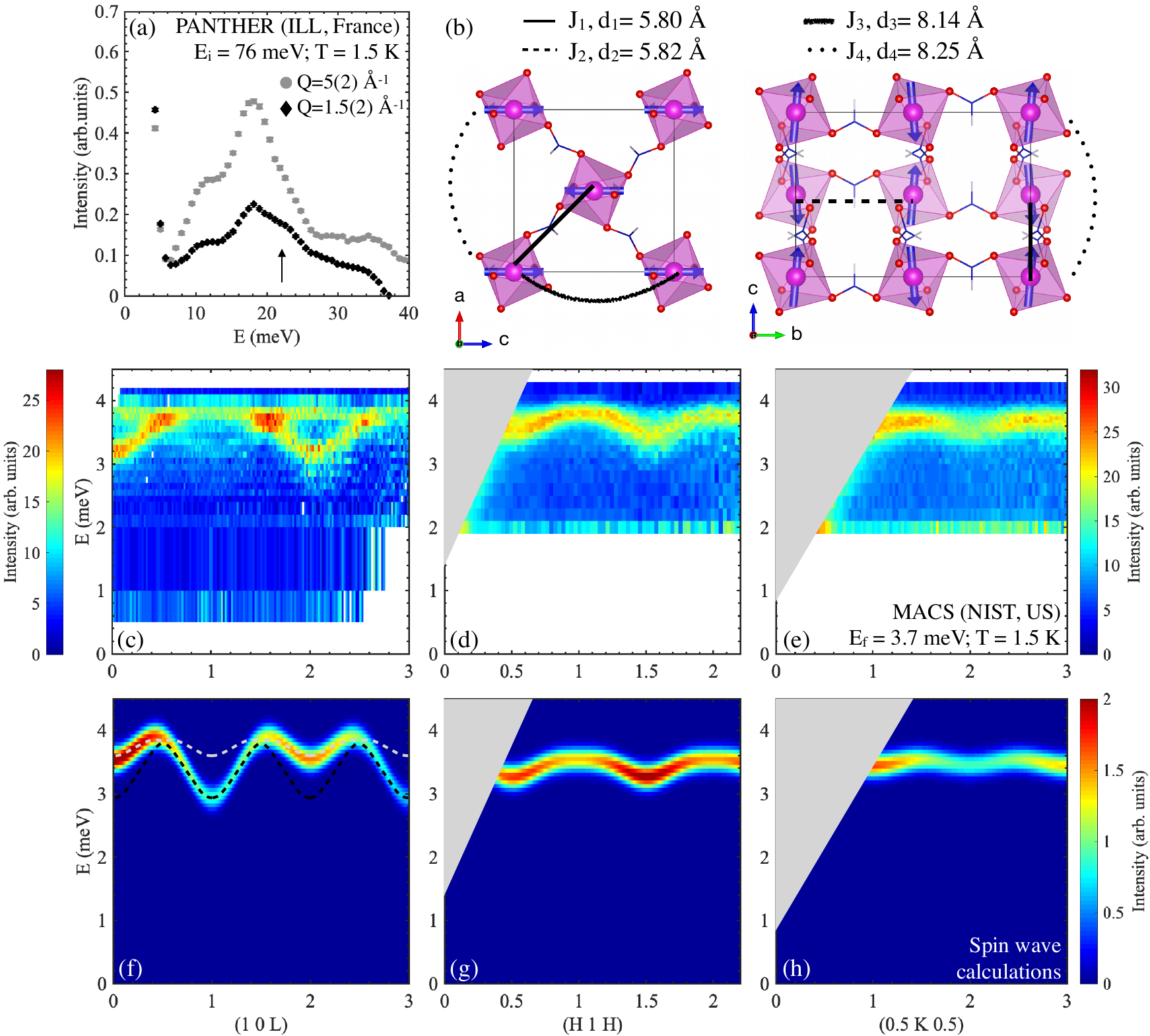}
\caption{(a) Constant-Q cuts (Q = 1.5$\pm$0.2 $\AA^{-1}$ Q = 5.0$\pm$0.2 $\AA^{-1}$) and extracted from the time-of-flight Panther data at 1.5~K and for the incident energy of E$_i$ = 76 meV. The black arrow indicates the excitation between the $j_{eff}$ = 1/2 and $j_{eff}$ = 3/2 manifolds due to spin-orbit coupling. (b) View of the magnetic structure of [CH$_3$NH$_3$][Co(HCOO)$_3$] and the exchange paths considered for the spin wave calculations as described in the text. (c-e) Neutron scattering intensity maps of the magnetic excitation spectra along the (10 L), (H 1 H) and (0.5 K 0.5) directions in reciprocal space at 1.5 K. The black stars represent the two magnon modes measured in THz spectroscopy. (f-h) Simulated magnetic excitation spectra using linear spin-wave theory. The dashed lines in panel (f) correspond to the calculated magnon dispersion relation, without taking into account the neutron cross-section.}\label{fig:macs}
\end{figure*}

In order to get further insight into the magnetic dynamical properties of MA-Co, we have measured the spin wave excitations, associated with the magnetic order using neutron spectroscopy. The magnetic excitation spectra measured on the cold spectrometer MACS (NIST, US) using two single crystals oriented in the (H~0~L) and the (H~K~H) scattering planes respectively, are presented in Fig. \ref{fig:macs} (c-e). The measured neutron spectra show dispersive branches along the  (1 0 L), (H 1 H) and (0.5 K 0.5) directions.  An energy gap of about 3 meV can also be observed in the three directions, which indicates a rather strong spin anisotropy. To model the spin dynamics, spin-wave calculations were performed using the SPINWAVE software \citep{SPINWAVE} developed at Laboratoire L\'eon Brillouin (LLB). In this system due to crystal field and spin-orbit coupling, the 3$d^7$ Co$^{2+}$ ions in an octahedral environment are in a high spin configuration $t^5_{2g}$$e^2_g$ with S = 3/2 and effective $\tilde{L}$ = 1 and form a $j_{eff}$ = 1/2 Kramers doublet ground state and $j_{eff}$ = 3/2 and 5/2 excited manifolds. Figure \ref{fig:macs} (a) shows constant-Q cuts at Q~=~1.5$\pm$0.2~$\AA^{-1}$ and Q~=~5$\pm$0.2~$\AA^{-1}$, extracted from the Panther data with E$_{i}$~=~76 meV at T~=~1.5~K. As the phonon and magnetic cross sections behave differently as a function of Q, it is possible to distinguish the spin-orbit excitation between the $j_{eff}$ = 1/2 and $j_{eff}$ = 3/2 manifolds from the phonon scattering by comparing theses cuts. Indeed, the phonon intensity increases with increasing Q while the magnetic form factor vanishes at large Q. By comparing the constant-Q cuts presented in Fig. \ref{fig:macs} (a), one can see an additional shoulder around 21~meV in the low-Q data. As its intensity vanishes on the high-Q data, we can assign this mode to the excitation between the  the $j_{eff}$ = 1/2 and $j_{eff}$ = 3/2 manifolds and extract a value of about 21 meV for the spin-orbit coupling. This value is consistent with other cobalt materials \cite{Wallington2015, Sarte2019, Songvilay2020}. On the other hand, the spin waves only extend up to 5 meV and are therefore well separated in energy from the excited $j_{eff}$ = 3/2 states, which validates the pseudospin $j_{eff}$ = 1/2 picture in this material.
In order to extract the values of the magnetic interactions in this compound, we chose to model the magnon dispersion using a Heisenberg Hamiltonian:
\[H = \mathop{\sum}_{i,j} J_{ij} \vec{S_i}\cdot \vec{S_j} + \mathop{\sum}_i D (S^{z}_{i} \cdot S^{z}_{i}) \]
with S = 1/2 spins, the last term denotes an easy-axis anisotropy. 
The linear spin wave calculations were performed taking into account the magnetic structure which was determined previously \cite{Mazzuca2018} and which is illustrated in Fig. \ref{fig:macs} (b).
Considering the distances between the Co$^{2+}$ cations, four magnetic exchange interactions are considered: $J_1$ (d~=~5.80 \AA) and $J_2$ (d~=~5.82\AA) represent the nearest- and next-nearest-neighbor interactions mediated by the formate molecules while $J_3$ and $J_{4}$ (d~=~8.14 and 8.25 \AA, respectively)  arise from a super-super exchange interaction between Co$^{2+}$ mediated by two formate molecules (Fig. \ref{fig:macs} b). 
The best agreement with the experimental data along the (1 0 L), (H 1 H) and (0.5 K 0.5) directions is shown in Figs. \ref{fig:macs} (f), (g) and (h), respectively. Note that the spin wave dispersion relation (without taking into account the neutron dynamical structure factor) contains two modes along the (1~0~L) direction ( represented as dashed lines in Fig. \ref{fig:macs} f), in agreement with the observations at $q=0$ in THz spectroscopy (Fig. \ref{fig:THz} a and c), at 3.1 meV and 3.8 meV. Due to a geometrical factor, the intensity varies between the two modes, at different points of the Brillouin zone in the calculation of the neutron cross-section in the (1 0 L) direction. On the other hand, the low statistics of the data in the (1 0 L) direction does not allow to confirm that the intensity around 3.8 meV arises from a second magnon mode that would be active at all points of the Brillouin zone. Moreover, this model shows calculated dispersions that are in close agreement with the data in the two other directions. The corresponding interactions are $J_1$ = 0.6(1) meV, $J_2$ = 0.8(1) meV, $J_3$ = -0.2(1) meV and $J_4$=0.3(1) meV. In addition, an easy-axis anisotropy term $D$ along the $c$-axis was adjusted to $D$ = 2.2 meV in order to account for the energy gap in the inelastic data, taking into account the fact that the spins lie along the $c$-axis in the orthorhombic symmetry, with a small canting along the $b$ axis. Note that this canting was attributed to a Dzyaloshinskii-Moriya interaction or single-ion anisotropy beyond first-order term \cite{Mazzuca2018}, which were not considered in our model due to its probable very small effects on the measured spin waves. Surprisingly, the third and fourth nearest-neighbour interactions are of the same order of magnitude as $J_1$ and $J_2$, and have to be of opposite signs (with $J_{4}$ antiferromagnetic) in order to properly reproduce the dispersions along the (1~0~L) and (H~1~H) directions (Fig. \ref{fig:macs} f-g)

\section{Discussion}

The combination of neutron powder diffraction, quasi-elastic neutron scattering, single crystal cold/thermal neutron spectroscopy and low-energy terahertz spectroscopy allows to span a large dynamical range and shed light on the microscopic mechanisms involved in the complex structural transitions of [CH$_3$NH$_3$][Co(HCOO)$_3$]. In a previous work, Canadillas-Delgado \textit{et al.} reported the detailed structural study of [CH$_3$NH$_3$][Co(HCOO)$_3$] throughout the different phase transitions, in particular the evolution of the hydrogen bonds as a function of temperature and their role in the stabilisation of the crystal structures \cite{Canadillas2019}. It is therefore interesting to compare their analysis of the static properties of [CH$_3$NH$_3$][Co(HCOO)$_3$] to our spectroscopy results, in order to obtain an overall view of the phase transitions in this material.

Starting from the room temperature orthorhombic phase (phase 1), Canadillas-Delgado \textit{et al.} reported a first change of state around 135~K, where the hydrogen atoms around the carbon atom of the CH$_{3}$ group of the methylammonium molecule are disordered and fluctuate over two crystallographic positions (Fig. \ref{fig:summary}, right panel), to a state where these hydrogen atoms order (phase 2). Although these atoms do not participate to any hydrogen bonding, their ordering is caused by a slight change of symmetry due to an anisotropic thermal contraction of the cell, as shown in our diffraction data. Indeed, in the room temperature phase, the CH$_{3}$NH$_{3}$ molecule sits in a mirror plane and there are 3 equivalent positions for the hydrogen atoms. Because of the thermal contraction of the cell, two out of three positions become independent, causing the hydrogen atoms to order. Although Canadillas-Delgado \textit{et al.} reported this phase transition around 135~K \cite{Canadillas2019}, our neutron diffraction, neutron inelastic and quasi-elastic data showed a change of behaviour around T$^{*}$~=~220~K. Above this temperature, the high energy harmonic phonons become smeared out and an additional fast dynamic component, that may be attributed to fast motions of hydrogen atoms, and which falls out of the instrumental energy window could be observed in the quasi-elastic data. The change of symmetry also affects the low-energy phonons observed in the THz range, with the splitting of the P1' mode (Fig. \ref{fig:THz} c-d). 

Meanwhile, above T$_{S2}$ and up to room temperature, the hydrogen atoms linked to the nitrogen atom in the NH$_{3}$ group of the CH$_{3}$NH$_{3}$ molecule establish two hydrogen bonds which are depicted in grey dashed lines in Fig. \ref{fig:summary}, while the third hydrogen bond still fluctuates. Below T$_{S2}$, Canadillas-Delgado \textit{et al.}  reported a transition from the commensurate orthorhombic (\textit{Pnma}) phase to a modulated \textit{Pnma} phase (phase 3), marked by the freezing of the third hydrogen atom around the nitrogen atom in the NH$_{3}$ group. This hydrogen atom now establishes two different hydrogen bonds with the same formate group, whose lengths vary as a function of temperature (depicted in dashed blue and green lines in Fig. \ref{fig:summary}). In our spectroscopy data, this transition marks the onset of a dynamical component in the quasi-elastic data, associated with the rotational dynamics of the whole CH$_{3}$NH$_{3}$ molecule: above T$_{S2}$, the molecule is free to fluctuate while the freezing of the third hydrogen bond around the NH$_{3}$ group below T$_{S2}$ marks the freezing of the molecule. Moreover, below this temperature, an additional phonon mode (P2) appears in the THz spectroscopy data. In this lowest energy range for the optical phonons (around 40 cm$^{-1}$ or 5 meV), the atomic vibrations are associated with the tilting of the inorganic framework, as described in previous studies on organic-inorganic halide perovskites  \cite{Swainson2015,Songvilay2019,Leguy2016,Glaser2015,Brivio2015}. The appearance of the P2 phonon mode below T$_{S2}$, the freezing temperature of the molecule, therefore attests of the coupling between the [CH$_{3}$NH$_{3}$] dynamics and the dynamics of the Co(COOH)$_{6}$ framework: above T$_{S2}$, it is likely to become anharmonic due to the strong molecular fluctuations.

The last transition is a structural transition from the orthorhombic to monoclinic symmetry below TS$_{1}$ (phase 4). In this low temperature phase, the six hydrogen positions around the nitrogen and carbon atoms of the CH$_{3}$NH$_{3}$ molecule become independent. In particular, around the nitrogen atoms, the hydrogen atoms now establish three fixed hydrogen bonds as shown in Fig. \ref{fig:summary} \cite{Canadillas2019}, which induces a slight re-orientation of the molecule, according to \cite{Canadillas2019, Mazzuca2018}. Interestingly, Mazzuca \textit{et al.} reported a large anomaly in the dielectric constant around this temperature and they interpreted this transition to be associated with an antiferroelectric-like order and later, Canadillas-Delgado \textit{et al.} interpreted this anomaly as a signature of the order-disorder transition of the methylammonium molecule. In contrast, our detailed dynamical study indicates rather that the order-disorder transition of the molecular cation actually occurs in two steps: first at T$_{S2}$, with the onset of the quasi-elastic neutron intensity and the emergence of the broad P'1 THz phononic mode, and secondly at T$_{S1}$ when P'1 becomes narrower, signature of collective, long-lasting vibrations. Moreover, the softening of the P2 phonon mode, associated with the inorganic framework, when approaching T$_{S1}$ as observed in THz spectroscopy, seems to also indicate a displacive-like character. 

Finally, our spin wave analysis in the magnetically ordered phase below T$_{N}$ confirms the presence of a large easy-axis anisotropy \cite{Mazzuca2018}, and allowed to extract magnetic exchange interactions of the order of 1 meV. It is interesting to note however that no particular change in the lattice and molecular dynamics was observed at the magnetic transition, therefore discarding any strong magneto-electric effect in this compound \cite{Claudia2016}. However, the presence of magnons and optical phonons separated by only 1 meV in energy in the THz range (less than 10 cm$^{-1}$) suggests that external parameters such as strong magnetic field or pressure may be relevant to generate noticeable a magneto-electric coupling.

\section{Conclusions}

In conclusion, we have investigated the spin, lattice and molecular dynamics of [CH$_3$NH$_3$][Co(HCOO)$_3$] using quasielastic, inelastic neutron scattering and THz spectroscopy techniques. The combination of these different techniques allowed to probe a large dynamical range and get further insight into the microscopic mechanisms involved in this complex material. In particular, the temperature dependence study of the optical phonon modes allowed to highlight the changes in the molecular dynamics associated with each structural phase transitions that were reported previously: the lattice distortion at T$^{*}$ corresponds to a first ordering of the hydrogen atoms around the carbon atom of the CH$_3$ methyl group. Then, the transition to the orthorhombic modulated phase below T$_{S2}$ is associated with an order-disorder transition of the whole CH$_3$NH$_3$ molecule. Finally, the  structural transition from orthorhombic to monoclinic symmetry at T$_{S1}$ is also of order-disorder type, that may be associated with the tilting of the Co(HCOO)$_3$ framework. The magnetic excitations were also mapped out using inelastic neutron scattering and analyzed using linear spin wave calculations. Overall, our THz and neutron spectroscopy data did not evidence any coupling between the magnetic and lattice degrees of freedom, indicating only very weak magneto-electric coupling in this compound, that could be amplified by external parameters such as magnetic field or pressure.

\acknowledgments
We acknowledge funding from ANR-18-CE09-0032-02 MONAFER, EPSRC and STFC. 
We thank J. Debray for the orientation of the single crystals and J. Robert for helpful discussions. We acknowledge the support of the National Institute of Standards and Technology, U.S. Department of Commerce, and ISIS neutron and muon source. Access to MACS was provided by the Center for High Resolution Neutron Scattering, a partnership between the National Institute of Standards and Technology and the National Science Foundation under Agreement No. DMR-1508249. We acknowledge the support of 2FDN for the access to D1B and SOLEIL for the access to the AILES beamline (proposal 20190483). 
%\bibliography{refMOF_MS}

%

\end{document}